\documentstyle[aps,pra,epsf,multicol]{revtex}
\begin{document}
\draft
\title{Proposal for measurement of harmonic oscillator Berry phase in ion traps}
\author{I. Fuentes-Guridi$^{*}$, S. Bose$^{\dagger}$  and  V. Vedral$^{\dagger}$}
\address{$^{*}$Optics Section, The Blackett Laboratory,
Imperial College, London SW7 2BZ, United Kingdom \\
$^{\dagger}$ Centre for Quantum Computation, Clarendon Laboratory,
    University of Oxford,
    Parks Road,
    Oxford OX1 3PU, England}

\maketitle
\begin{abstract}
We propose a scheme for measuring the Berry phase in the vibrational degree
of freedom of a trapped ion. Starting from the ion in a vibrational
coherent state we show how to reverse the sign of the coherent state amplitude
by using a purely geometric phase. This can then be detected through the
internal degrees of freedom of the ion. Our method can be applied
to preparation
of Schr\"odinger cat states.
\end{abstract}


\begin{multicols}{2}
When the Hamiltonian of a quantum system is varied adiabatically in a cyclic
fashion, the state of the system acquires a geometrical phase in addition to
the usual dynamical phase. This effect, discovered by Berry \cite{berry},
(and generalized in various ways \cite{aa}) has been widely 
tested for $2$-state systems \cite{tom}
and attracted interest from a variety of fields \cite{wil,oth}.
However, the Berry phase has not been experimentally measured for quantum harmonic
oscillators, though some theoretical calculations exist \cite{chatur,qh,chiao,ger,pati}. 
The reason for this might be the fact that for the simplest case, namely
for adiabatic displacement of an oscillator state in phase
space, the Berry phase is independent of the state \cite{chatur}, and thereby undetectable.
However, when a squeezing Hamiltonian is switched on, and the squeezing
parameter is varied, there would be a detectable
Berry phase \cite{chatur,qh,chiao}. For an initial Fock state $|n\rangle$ which
undergoes squeezing, the Berry phase after a cycle is $-(n+1/2)$ times the
classical Hannay angle \cite{chatur,qh}. The squeezed states of the
electromagnetic field would have been a natural candidate to test this kind of
phase, but they are not stable enough for an
adiabatic evolution. However, the vibrational mode of a trapped ion has been a
fertile ground for the preparation of long lived nonclassical states of a
harmonic oscillator \cite{thom,hein,coh,Scat,stein}. In this letter, we derive
a Berry phase formula for a certain adiabatic evolution of a joint state of
the internal levels of a trapped ion and its vibrational motion. Despite being
the phase gained by a joint state, its value is {\em fundamentally dependent} on the
harmonic oscillator nature of the vibrational mode. We propose a scheme to
detect this phase which is feasible with current technology. It is worthwhile
to mention that based on calculations of Ref.\cite{chiao} (which where
tested in systems other than quantum harmonic oscillators \cite{dim}), there has been an
earlier attempt to detect the {\em nonadiabatic} geometric phase in ion traps by applying a set of four squeezes to the vibrational state \cite{king}. Here we  
propose a way of detecting the harmonic oscillator version of Berry's original
{\em adiabatic} geometric phase. 

 Consider the Hamiltonian
\begin{equation} \label{eq:ham}
H=H_{a}+H_{b},
\end{equation}
where
\begin{equation} \label{eq:hama}
H_{a}=g_{a}e^{i\phi}\vert e\rangle\langle g\vert a+h.c.,
\end{equation}
\begin{equation} \label{eq:hamb}
H_{b}=g_{b}\vert e\rangle\langle g\vert a^{\dagger}+h.c.\;\; .
\end{equation}
In the above,
$|e\rangle$ and $|g\rangle$ are two states of a qubit, $a^{\dagger}$ and $a$
are the creation and annihilation
operators of a harmonic oscillator, $g_{a}$ and $g_{b}$ are unequal positive
interaction strengths (say $g_{a}>g_{b}$) and $\phi$ is
an arbitrary phase factor. The motivation for choosing this Hamiltonian is
the possibility of its physical implementation and this will be
described later. If the phase $\phi$ is slowly varied over a complete loop (so that the adiabatic approximation holds true), there will
be a nontrivial Berry phase acquired by an eigenstate of the Hamiltonian $H$.
We now proceed to calculate this.
We transform the Hamiltonian as
\begin{equation}
H^{'}=S(\epsilon)^{\dagger} H S(\epsilon)
\end{equation}
where
$S(\epsilon)^{\dagger}aS(\epsilon)=a\cosh(r)-a^{\dagger}\sinh(r)e^{i\theta}$
is a squeezing transformation with squeezing parameter $\epsilon=r e^{i \theta}$. If we chose the squeezing strength $r=\tanh^{-1} g_{b}/g_{a}$
and the squeezing phase $\theta=-\phi$, the transformed Hamiltonian will
be
\begin{equation}
H^{'}=\Omega(\vert e\rangle\langle g\vert a+ a^{\dagger}
\vert g\rangle \langle e\vert),
\end{equation}
where $\Omega= g_{a}\cosh(r)-g_{b}\sinh(r)$.
$H^{'}$ is the well known resonant Jaynes-Cummings Hamiltonian \cite{knight}.
A similar transformation to arrive at a
Jaynes Cummings Hamiltonian has been
used in Refs.\cite{dark,cirac}. The eigenstates of $H^{'}$ are
\begin{equation}
\vert \Psi_{n}^{\pm}\rangle=\frac{1}{\sqrt{2}}(\vert g\rangle\vert n+1\rangle\pm\vert e\rangle\vert n\rangle).
\end{equation}
This implies that the eigenstates of our original Hamiltonian $H$ are
\begin{equation}
\label{eig}
\vert \Psi_{n}^{\pm}(\epsilon) \rangle=\frac{S(\epsilon)}{\sqrt{2}}(\vert g\rangle\vert
n+1\rangle\pm\vert e\rangle\vert n\rangle).
\end{equation}
The states $S(\epsilon)\vert n\rangle$ featuring in the above expression
are the squeezed number states \cite{knight1}. 
Now we can proceed to calculate the Berry phase from the instantaneous eigenstates $\vert \Psi_{n}^{\pm}(\epsilon) \rangle$ of $H$.

  The expression for the Berry phase for an adiabatic cyclic evolution of
a Hamiltonian $H(\textbf{R})$ in parameter space $\textbf{R}$ is given in terms of the
instantaneous eigenstates $\vert n,\textbf{R}\rangle$ of the Hamiltonian
as
\begin{equation}
\gamma_{n}=i\int_{c}d\textbf{R}\langle n,\textbf{R}\vert\nabla_{R}\vert n,\textbf{R}\rangle.
\end{equation}
Using Eq.(\ref{eig}) in the above equation we get

\begin{eqnarray}
\gamma_{n}&=&i\int_{c}d\epsilon
\langle \Psi_{n}^{\pm}\vert S(\epsilon)^{\dagger}\nabla_{\epsilon}S(\epsilon)
\vert \Psi_{n}^{\pm}\rangle \nonumber\\
&=& \frac{i}{2}\int_{c}d\epsilon
\langle n+1\vert S(\epsilon)^{\dagger}\nabla_{\epsilon}S(\epsilon)
\vert n+1\rangle \nonumber \\&+&\frac{i}{2}\int_{c}d\epsilon
\langle n\vert S(\epsilon)^{\dagger}\nabla_{\epsilon}S(\epsilon)
\vert n\rangle.
\end{eqnarray}
If the modulus $r$ of the parameter $\epsilon$ is kept
constant throughout the evolution, then using the expression for
$\langle n\vert S(\epsilon)^{\dagger}\nabla_{\epsilon}S(\epsilon)
\vert n\rangle$ from Ref.\cite{chatur} we get
\begin{equation}
\label{bp}
\gamma_{n}=-2\pi(n+1)\sinh^{2}{r}.
\end{equation}
Note that the Berry phase $\gamma_{n}$ is same for both the states
$\vert \Psi_{n}^{+}(\epsilon) \rangle$ and $\vert \Psi_{n}^{-}(\epsilon) \rangle$. However, the dynamical phase $\beta_n^{\pm}$ is exactly opposite
for the eigenstates $\vert \Psi_{n}^{+}(\epsilon) \rangle$ and $\vert \Psi_{n}^{-}(\epsilon) \rangle$
and is
given by
\begin{eqnarray}
\beta_n^{\pm}&=& -\int_{c}dt
\langle \Psi_{n}^{\pm}\vert S(\epsilon)^{\dagger} H S(\epsilon)
\vert \Psi_{n}^{\pm}\rangle \nonumber \\
&=& \mp \Omega \sqrt{n+1} T,
\end{eqnarray}
where $T$ is the time period of the cyclic variation of $\epsilon$.
Thus one can make the dynamical phase completely vanish after two cycles
by changing
the state $\vert \Psi_{n}^{+}(\epsilon) \rangle$ to $\vert \Psi_{n}^{-}(\epsilon) \rangle$ or vice versa after one cycle. Under such
circumstances, the only contribution to the phase of the system will be
geometrical.

  Let us now describe how the Hamiltonian $H$ of Eq.(\ref{eq:ham}) can be
physically realized. Recently, ion traps have been
a very active field of both experimental \cite{thom,hein,coh,Scat} and theoretical \cite{stein,dark,cirac,zol1} research. Consider a single $\Lambda$
three-level ion with two hyperfine ground states and an excited state
(such as $^9 Be^{+}$ \cite{thom}) in a harmonic trap of frequency $\nu$ .  The two ground states, labelled as $|e\rangle$
and $|g\rangle$, are
separated in frequency by an amount $\omega_{o}$ which is much less
than their separation from the excited state. The motion of the ion is modified by its interaction with two pairs of travelling-wave laser beams whose frequencies
are detuned from the excited state. The first pair of lasers have frequencies $\omega_{L1}$ and $\omega_{L2}$ which satisfy $\omega_{L1}-\omega_{L2}=\omega_{o}+\nu$ and the second pair of lasers have frequencies $\omega_{L3}$ and $\omega_{L4}$ which satisfy $\omega_{L3}-\omega_{L4}=\omega_{o}-\nu$. We require that the pair
of lasers $L1$ and $L2$ differ in frequency from the pair $L3$ and $L4$ by
by an amount much larger than $\omega_0$. We also assume that $L1$
differs in phase from the rest of the beams by an amount $\phi$ (this phase will need to be slowly varied). 
The Hamiltonian for this system in the rotating frame with $U=\exp(-i\omega_{0}\sigma_{z}t)$ and after making the optical rotating
wave approximation is 
\begin{eqnarray}
\label{eq13}
H^{(1)}&=&\Omega_{12}(e^{i(\phi-\nu t+\eta_{12} (a +a^{\dagger}))}\sigma_{+}+ h.c) \nonumber \\&+&
\Omega_{34}(e^{i(\nu t+\eta_{34} (a +a^{\dagger}))}\sigma_{+}+ h.c)
\end{eqnarray}
where $a^{\dagger}$ and $a$
are the creation and annihilation
operators for the vibrational modes, $\sigma_{-}=|g\rangle\langle e|$, $\sigma_{+}=|e\rangle\langle g|$ and $\sigma_{z}=|g\rangle\langle g|-
|e\rangle\langle e|$, $\Omega_{ij}$ are Rabi frequencies of
$|e\rangle \rightarrow |g\rangle$ transition induced by the $i$th and 
$j$th lasers and the factor $\eta_{ij}= \delta k_{ij} a_{o}$ is the
Lamb-Dicke parameter (where $a_{o}$ is the amplitude of the ground state of the trap potential
and $\delta k_{ij}$ is the wave vector difference between the $i$th and the
$j$th beams). Hamiltonians comprising of 
any one of terms $\Omega_{12}(e^{i(\phi-\nu t+\eta_{12} (a +a^{\dagger}))}\sigma_{+}+ h.c)$ or $\Omega_{34}(e^{i(\nu t+\eta_{34} (a +a^{\dagger}))}\sigma_{+}+ h.c)$ have
been used in the context of nonclassical state preparation of the vibrational
mode \cite{thom} and in a recent experiment such terms where switched on
simultaneously \cite{natw}. These are generally implemented with a single pair of travelling
wave laser beams. We require two pairs of laser beams to switch on both 
the Hamiltonian terms simultaneously.
Similar Hamiltonian terms, when used simultaneously and in
conjunction with atomic decay, can be used to generate squeezed states of motion of the ion \cite{dark,cirac}. Here, however, we do not
want atomic spontaneous emission, and thus choose the $|e\rangle$ and $|g\rangle$ to be
hyperfine levels so that
 atomic decay can be neglected. 
In the Lamb-Dicke limit ($\eta_{ij}<<1$)  the field can be expanded to the first
order in $\eta_{ij}$. Expanding thus, and
transforming into the rotating frame with $U=\exp(-i\nu a^{\dagger}a t)$
we obtain
\begin{eqnarray}
H^{(1)}=U^{\dagger}HU&=&
|e\rangle \langle g|(g_{a}e^{i\phi}a+g_{b}a^{\dagger})+h.c. \nonumber \\
&=&H_a+H_b
\end{eqnarray}
where we have made another rotating wave approximation by omitting all the
 rapidly oscillating terms. In the above, $g_{a}=\eta_{12}\Omega_{12}/2$
and  $g_{b}=\eta_{34}\Omega_{34}/2$. Thus we can
obtain the Hamiltonian $H$ in an ion trap by the above methods in a rotating
frame. From an experimental perspective,
instead of slowly varying the phase of one of the beams relative to the
rest, one can alternatively keep the phase of all lasers
same but set $\omega_{L1}-\omega_{L2}=\omega_{o}+\nu+\delta$ where $\delta << \nu$. Then
$t\delta$ which varies very slowly with time $t$ will serve as the varying phase
$\phi$. There is also an alternative way to implement the
required Hamiltonian which decreases the number
of laser beams to two. In this case, each laser drives a quadrupole transition
such as the narrow $S_{1/2}\rightarrow D_{5/2}$ transition in $^{40} Ca^{+}$
\cite{na}
and the lasers would have to be oppositely detuned from the transition by $\nu$.

       We now describe our proposal to measure a Harmonic oscillator
Berry phase in a
vibrational mode of a trapped ion. It is assumed that the atom is cooled
to the ground state of motion and in the internal state $|g\rangle$.
First we prepare a coherent state $|\alpha \rangle$ of the vibrational
mode. This can be done by shifting the centre of the trap \cite{hein}
or other methods \cite{dark,coh} (our scheme is independent of
the method used). Next we want to achieve the evolution
\begin{equation}
|g\rangle|\alpha\rangle \rightarrow |g\rangle|-\alpha\rangle
\end{equation}
by purely geometrically. To this end we switch on the Hamiltonian $H$ by
switching on appropriate lasers and viewing the dynamics from the rotating
frame with $U=\exp[-i(\nu a^{\dagger}a+\omega_{0}\sigma_{z})t]$ (we will describe later what happens in the laboratory frame). The initial state can be written as
\begin{equation}
|\alpha\rangle|g\rangle=\frac{e^{-|\alpha|^2/2}}{\sqrt{2}}\sum_{n}\frac{\alpha^{n}}{\sqrt{n!}}
(\vert \Psi_{n}^{+}\rangle + \vert \Psi_{n}^{-}\rangle).
\end{equation}
Next, the relative phase $\phi$ of the lasers is varied very slowly and
cyclically while the parameter $r=\tanh^{-1} g_{b}/g_{a}$ is kept
constant. If $\phi$ is varied slowly enough so that adiabaticity holds,
the state evolves as
\begin{eqnarray}
|\Psi(\phi)\rangle &=&\frac{e^{-|\alpha|^2/2}}{\sqrt{2}}\sum_{n}\frac{\alpha^{n}}{\sqrt{n!}}
e^{i \gamma_{n}(\phi)} (e^{i\beta_n^{+}(\phi)}\vert \Psi_{n}^{+}(\epsilon) \rangle \nonumber \\&+&
e^{i\beta_n^{-}(\phi)}\vert \Psi_{n}^{-}(\epsilon) \rangle),
\end{eqnarray}
where $\gamma_{n}(\phi)$ and  $\beta_n^{\pm}(\phi)$ are
geometric  and dynamical phases respectively and $\epsilon=r e^{-i\phi}$. After $\phi$ has completed an entire cycle (i.e. $\phi=2\pi$),
we apply a state dependent phase shift $|g\rangle \rightarrow -|g\rangle$ to the ionic state (by applying a $2\pi$ pulse \cite{Scat}) and this has to be done  much faster than the evolution timescale of the system. This converts $\vert \Psi_{n}^{+}(\epsilon) \rangle \rightarrow \vert \Psi_{n}^{-}(\epsilon) \rangle$ and vice versa. Now we vary the phase difference
$\phi$ again from $2\pi$ to $4\pi$. The resulting state at the end of this second cycle
is
\begin{eqnarray}
|\Psi(4\pi)\rangle &=&
\frac{e^{-\frac{|\alpha|^2}{2}}}{\sqrt{2}}\sum_{n}\frac{\alpha^{n}e^{i (\gamma_{n}(4\pi)+\beta_n^{+}(2\pi)+\beta_n^{-}(2\pi))}}{\sqrt{n!}}
(\vert
\Psi_{n}^{+}(\epsilon) \rangle \nonumber \\&+& \vert
\Psi_{n}^{-}(\epsilon) \rangle).
\end{eqnarray}
As $\beta_n^{+}=- \beta_n^{-}$, the dynamical phase completely cancels. If
we choose $\sinh^{2}{r}=1/4$, we have $\gamma_{n}(4\pi)=-n\pi$.
Under such circumstances, the final state would be
\begin{eqnarray}
|\Psi(4\pi)\rangle &=& \frac{e^{-|\alpha|^2/2}}{\sqrt{2}}\sum_{n}\frac{(-\alpha^{n})}{\sqrt{n!}}
 (\vert \Psi_{n}^{+}(\epsilon) \rangle +
\vert \Psi_{n}^{-}(\epsilon) \rangle) \nonumber \\
&=& |g\rangle|-\alpha\rangle.
\end{eqnarray}
So the detection of the Berry phase now amounts to distinguishing
between the coherent states $|\alpha\rangle$ and
$|-\alpha\rangle$. To this end, after switching off the adiabatic
evolution, the entire state is given a negative displacement of
$-\alpha$. This reduces our problem to distinguishing
between $|0\rangle$ (a vibrational state with no excitation)
and $|-2\alpha\rangle$ \cite{har}. To accomplish this, the ionic internal states are allowed to
interact with the vibrational mode by a Jaynes Cummings
interaction \cite{har}. In the case of no Berry phase, the probability of finding
the ion in an excited state at any time $t$ is 
$P_{eo}=0$, while, in the
presence of a Berry phase the same probability is given by
\begin{equation}
P_{e,-2\alpha}=e^{-2|\alpha|^2}\sum \frac{(-2\alpha)^{2n}}{n!}
\sin^2{\frac{\Omega_{n+1}}{2}t}
\end{equation}
where $\Omega_{n+1}$ is the Rabi frequency corresponding to an
excitation number $n+1$. Note that the above method is not a perfect discrimination
(it is impossible in principle as $|0\rangle$ and
$|-2\alpha\rangle$ are nonorthogonal).  
Incidentally, the squeezing in our proposal ($r=0.48$)
is smaller than the value $r\sim 1.5$ required for the four squeeze nonadiabatic 
geometric phase proposal \cite{king}. If the original coherent
state amplitude is $|\alpha| \sim 1$, we can ensure that this squeezing
does not bring in 
contribution from the higher order terms
in the expansion of Eq.(\ref{eq13}) (From Ref.\cite{knight1}
one can show that the probability of the state $|3\rangle$ in squeezing
of $|1\rangle$ with $r\sim 0.48$ is already small enough to make the ratio
of higher order terms to the first order term lower than $10^{-2}\eta^2$). One alternative to
using a coherent state is to start with the superposition $|0\rangle+|1\rangle$
(which should not be too difficult to prepare \cite{pvking}) and convert it to the state $|0\rangle-|1\rangle$ by following an identical procedure to that
described above for coherent states.  
   
  In the laboratory (nonrotating) frame, there will be an extra phase
equal to $\int_{c}d\epsilon
\langle \Psi_{n}^{\pm}\vert S(\epsilon)^{\dagger} (\nu a^{\dagger}a+\omega_{0}\sigma_{z})  S(\epsilon)
\vert \Psi_{n}^{\pm}\rangle$. For our choice of $\sinh^{2}{r}=1/4$ this phase
 turns out to be $3 \nu T$ where $T$ is the time required 
to complete {\em one} cycle of the phase $\phi$. We can choose $T$ in such
a way that this phase becomes a multiple of $2\pi$. This will keep the
magnitude of the phase unaltered from that in the rotating frame.      

   Let us now examine the feasibility of our experiment with existing
ion trap parameters. For adiabaticity, we require the time scale $T$
of variation of the
relative phase $\phi$ to be greater than the dynamical timescale of
the problem. The dynamical timescale is given by 
$g_{a/b}^{-1}=(\Omega_{ij} \eta_{ij}/2)^{-1}$. We first consider
experiments with $^9 Be^{+}$ ion \cite{thom,natw} where
$\eta \sim 0.2$  and $\Omega/2\pi \sim 500$ kHz. The 
dynamical timescale, required to be smaller than $T$, is then about $0.33\times10^{-5}$ s. $T$, on the other hand, is required to be lower than both the lifetime of the excited state (which can be up to $10$ s \cite{cirac}) and the
motional decoherence timescale (which is about $10^{-4}$ s \cite{thom} for $^9 Be^{+}$ experiments) so
that our assumption of neglecting all types of decoherence holds. If we set $T\sim 10^{-5}$ s (much less than both the decoherence
time scales), then it is about thrice the dynamical
timescale and the assumption of adiabaticity should hold (one could even try to ensure
an order of magnitude difference between the dynamical timescale and
 by increasing the laser
power a bit more than three folds). On the other hand, motional decoherence, which is an
order of magnitude slower than $T$, can be neglected. 
In $^{40} Ca^{+}$ experiments \cite{ros}, the dynamical timescale is 
$10^{-5}$ s whereas the motional decoherence time scale is $10^{-3}$ s
(internal state decoherence being $1$ s). For these traps,
we can choose $T\sim 10^{-4}$ s and perfectly satisfy both adiabaticity and
the neglect of decoherence. We also have to set $3\nu T = m 2\pi$ for the extra phase in the nonrotating
frame to vanish, where $m$ is an integer. For standard traps, $\nu \sim 10$ MHz,
\cite{dark} and hence $\nu T \sim 10^4$. We can easily choose a number of such a
large order to be a multiple of $2\pi$.

 It is interesting to point out that if our scheme is performed starting with the ion in
the excited state, the evolution would be
$|e\rangle |\alpha\rangle \rightarrow -|e\rangle |-\alpha\rangle$.
Therefore, the initial 
internal state of the ion makes no difference to our scheme. Using a similar
scheme we can also create a Schr\"odinger cat state of the ion. One initially
has to prepare the ion in a superposition of $|g\rangle$ and some
other state $|r\rangle$ which is completely decoupled from the evolution
due to our Hamiltonian. Then the evolution proceeds as 
\begin{equation}
\label{scat}
(|g\rangle + |r\rangle)|\alpha\rangle \longrightarrow |g\rangle |-\alpha\rangle+ |r\rangle|\alpha\rangle\;. 
\end{equation}
The state in the above equation is a Schr\"odinger cat state of the ion \cite{Scat}.
    
  We have shown how to observe the Berry phase of a simple harmonic
oscillator using a trapped ion. We have described how to
reverse the amplitude of a coherent state by purely geometric means (i.e. by
using only the Berry phase). An important advantage of the experiment is that it needs only a single trapped ion.
This is a simple requirement in comparison to the technology that
already exists such as the ability to entangle four ions in a trap
\cite{natw}. Moreover, we have shown that the existing ion trap parameters are well in range of those
required for implementing our proposal and for our analysis to remain
valid. We have also indicated how to use our scheme to create a Schr\"odinger cat
state of the ion. Further interesting extensions to
the geometric approach to multiple modes are also possible and will be
investigated in the future.

We thank P. L. Knight and B. E. King for a very careful reading of the manuscript and valuable
comments. 
We would also like to thank J. I. Cirac, R. C. Thompson, J. Pachos and A. K. Pati for 
discussions. This research has been partly supported by
the European Union, EPSRC and Hewlett-Packard. I. F.-G. would like to thank Consejo Nacional de Ciencia y Tecnologia (Mexico) Grant no. 115569/135963 for financial support.

\end{multicols}
\end{document}